# ANAIS–112: updated results on annual modulation with three-year exposure

**Iván Coarasa,**[a,b,*] **Julio Amaré,**[a,b] **Jaime Apilluelo,**[a,b] **Susana Cebrián,**[a,b] **David Cintas,**[a,b] **Eduardo García,**[a,b] **María Martínez,**[a,b] **Miguel Ángel Oliván,**[a,b,c] **Ysrael Ortigoza,**[a,b,d] **Alfonso Ortiz de Solórzano,**[a,b] **Tamara Pardo,**[a,b] **Jorge Puimedón,**[a,b] **Ana Salinas,**[a,b] **María Luisa Sarsa**[a,b] **and Patricia Villar**[a]

[a]*Centro de Astropartículas y Física de Altas Energías (CAPA), Universidad de Zaragoza,*
 *Pedro Cerbuna 12, 50009 Zaragoza, Spain*

[b]*Laboratorio Subterráneo de Canfranc,*
 *Paseo de los Ayerbe s.n., 22880 Canfranc Estación, Huesca, Spain*

[c]*Fundación CIRCE,*
 *Av. de Ranillas 3D, 50018 Zaragoza, Spain*

[d]*Escuela Universitaria Politécnica de La Almunia de Doña Godina (EUPLA), Universidad de Zaragoza,*
 *Calle Mayor 5, La Almunia de Doña Godina, 50100 Zaragoza, Spain*

 *E-mail:* icoarasa@unizar.es

The ANAIS experiment is intended to search for dark matter annual modulation with ultrapure NaI(Tl) scintillators in order to provide a model independent confirmation or refutation of the long-standing DAMA/LIBRA positive annual modulation signal in the low energy detection rate, using the same target and technique. Other experiments exclude the region of parameters singled out by DAMA/LIBRA. However, these experiments use different target materials, so the comparison of their results depends on the models assumed for the dark matter particle and its distribution in the galactic halo. ANAIS–112, consisting of nine 12.5 kg NaI(Tl) modules produced by Alpha Spectra Inc., disposed in a 3×3 matrix configuration, is taking data smoothly with excellent performance at the Canfranc Underground Laboratory, Spain, since August, 2017. Last published results corresponding to three-year exposure were compatible with the absence of modulation and incompatible with DAMA/LIBRA for a sensitivity above $2.5\sigma$ C.L. Present status of the experiment and a reanalysis of the first 3 years data using new filtering protocols based on machine-learning techniques are reported. This reanalysis allows to improve the sensitivity previously achieved for the DAMA/LIBRA signal. Updated sensitivity prospects are also presented: with the improved filtering, testing the DAMA/LIBRA signal at $5\sigma$ will be within reach in 2025.

[*]Speaker

## 1. Introduction

ANAIS [1, 2] (Annual modulation with NaI(Tl) Scintillators) is a direct dark matter detection experiment aiming at the confirmation or refutation of the DAMA/LIBRA positive annual modulation signal in the low energy detection rate (1-6 keV) [3], using the same target and technique. This signal, observed for more than 20 years with a statistical significance close to $13\sigma$, is in strong tension with the negative results of other very sensitive experiments [4], but a direct comparison using the same target material was lacking.

ANAIS–112 is operating an array of 3×3 ultrapure NaI(Tl) crystals with a total mass of 112.5 kg at the Canfranc Underground Laboratory (LSC) in Spain since August 3, 2017. Six-year exposure (643.5 kg×y) has already been completed with excellent duty cycle, about 95% of live time. The main features of the ANAIS modules, built by Alpha Spectra Inc., include the incorporation of a Mylar window in the lateral face allowing low energy calibration using external gamma sources, and their exceptional optical quality, which added to the high efficiency Hamamatsu photomultipliers coupled to the crystals enable a light collection at the level of 15 photoelectrons/keV [1] in all the modules. All the nine modules are simultaneously calibrated with external $^{109}$Cd sources every two weeks to correct possible gain drifts. The calibration of the region of interest (ROI, 1-6 keV) is finally performed with the bulk $^{22}$Na (0.87 keV) and $^{40}$K (3.20 keV) events identified by coincidences with high energy depositions (1274.5 and 1460.8 keV, respectively) in a second module. The use of $^{22}$Na/$^{40}$K lines increases the reliability of the energy calibration in the ROI, since they are actually either in the ROI or very close to the energy threshold, set at 1 keV [1].

The trigger rate in the ROI is dominated by non-bulk scintillation events. For this reason, the development of robust protocols for the selection of events corresponding to bulk scintillation in sodium iodide is mandatory. The selection criteria initially developed and applied in ANAIS–112 have been exhaustively described in [1] and they are based on standard cuts on a few parameters. Although this filtering procedure works very well above 2 keV, in the region from 1 to 2 keV it shows some weaknesses. On the one hand, the efficiency of acceptance of bulk scintillation events decreases very sharply, reducing the sensitivity, and on the other hand, the measured rate in this region is about 50% higher than expected according to the ANAIS–112 background model [5], pointing at a possible leaking of non-bulk scintillation events as responsible of that excess. In order to improve the rejection of noise events between 1 and 2 keV, a machine-learning technique based on a Boosted Decision Tree (BDT) has been implemented. The detailed description of the BDT performance in ANAIS–112 can be found in [6].

## 2. Machine-learning technique for event selection

As training populations for BDT, we combine [6]: as signal events, scintillation events in [1-2] keV in the bulk of the crystals produced by neutron interactions from dedicated neutron calibrations with a $^{252}$Cf source located outside the ANAIS–112 shielding, which are mostly associated to elastic nuclear recoils in that energy region [7]; and as noise events, those coming from a blank module similar to the ANAIS–112 modules, but without NaI(Tl) crystal. This choice of the training populations is robust, because it does not rely on background events at all, and uses bulk events as signal. As a result of the training, a new parameter is built, named BDT, that combining the information of the 15 discriminating parameters maximizes the discrimination between the signal and noise populations used in the training. For the event selection, we define an energy-dependent BDT cut and keep only events above it. This selection has been optimized for each detector and energy bin. The corresponding efficiency is estimated for each detector independently by using $^{252}$Cf neutron calibration events. The ratio of the events which pass the signal selection to the total events is the acceptance efficiency. The acceptance efficiency derived from the BDT cut is significantly higher (around 30% in [1-2] keV) with respect to the previous ANAIS–112 filtering procedure. Furthermore, the BDT method developed significantly reduces the background level below 2 keV for all detectors with respect to that obtained by the previous ANAIS–112 protocols. In particular, the integral rate from 1 to 2 keV is 5.39±0.04 and 4.40±0.03 c/keV/kg/d for the ANAIS–112 filtering procedure and the BDT method, respectively, which represents a reduction of the background of 18%.

## 3. Annual modulation reanalysis with three-year exposure

After the application of the BDT for event selection, we have performed the reanalysis of the first three years of ANAIS–112 data searching for annual modulation in the same regions as DAMA/LIBRA has published ([1-6] and [2-6] keV), using an exposure of 322.83 kg×y. Following a similar fitting strategy as in [2], a simultaneous fit of all the detectors' data with different background parameters, but the same modulation parameters, is carried out. The results of the $\chi^2$ minimization following Equation (6) in [2] are: the null hypothesis is well supported by the $\chi^2$ test in both energy regions, with $\chi^2$/NDF=993.8/972 (p-value=0.307) from 1 to 6 keV and $\chi^2$/NDF=958.6/972 (p-value=0.615) between 2 and 6 keV; the best fits for the modulation hypothesis are $S_m = -0.0033 \pm 0.0037$ c/keV/kg/d (p-value=0.305) and $0.0007 \pm 0.0037$ c/keV/kg/d (p-value=0.606) for [1-6] and [2-6] keV, respectively, both compatible with zero at $1\sigma$, and incompatible with DAMA/LIBRA result at $3.9\sigma$ and $2.8\sigma$, respectively. We quote our sensitivity to DAMA/LIBRA as the ratio of DAMA modulation result over the standard deviation on the modulation amplitude derived from ANAIS–112 data ($S = S_m^{\text{DAMA}}/\sigma(S_m)$). At present, the achieved sensitivity is at $2.9\sigma$ in both energy regions. As can be seen in Figure 1, these results (black dots) also confirm our sensitivity prospects updated with respect to those previously calculated [8] considering the background level after filtering with the BDT. In particular, the BDT method improves the sensitivity between 1 and 6 keV by about 10% with respect to that obtained with the previous ANAIS–112 filtering protocols. Moreover, $5\sigma$ sensitivity will be within reach by the end of 2025.

## 4. Conclusions

A new low-energy noise event filtering protocol based on the boosted decision tree technique has been developed for the ANAIS–112 experiment. With this filtering procedure, a background level reduction of around 20% has been achieved between 1 and 2 keV. However, an excess of events below 2 keV with respect to our background model is still present, pointing at a background contribution in this energy range unaccounted for in the model and supporting the convenience of the revision of the ANAIS–112 background model. Furthermore, the reanalysis of the first three years of ANAIS–112 data with this technique has been presented. We obtain for the best fit a modulation amplitude of $-0.0033 \pm 0.0037$ c/keV/kg/d ($0.0007 \pm 0.0037$ c/keV/kg/d) in the [1-6] keV ([2-6] keV) energy region, supporting the absence of modulation in our data, and being incompatible with the DAMA/LIBRA result at $3.9\sigma$ ($2.8\sigma$), with a present sensitivity of $2.9\sigma$. According with our sensitivity estimates, which are confirmed with the latter results, the ANAIS–112 experiment will reach $5\sigma$ sensitivity by the end of 2025.

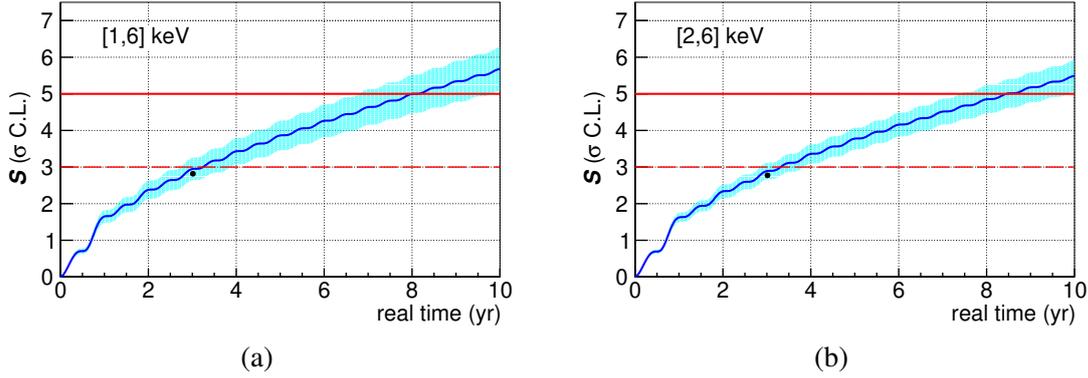

**Figure 1:** ANAIS–112 sensitivity to the DAMA/LIBRA signal recalculated using the background resulting after applying the BDT cut in $\sigma$ C.L. units as a function of real time in the [1-6] keV (a) and [2-6] keV (b) energy regions. The black dots are the sensitivities derived from the reanalysis of the three years of ANAIS–112 data using the BDT method. The cyan bands represent the 68% C.L. DAMA/LIBRA uncertainty.

## Acknowledgments

This work has been financially supported by MCIN/AEI/10.13039/501100011033 under grant PID2019-104374GB-I00, the Consolider-Ingenio 2010 Programme under grants MultiDark CSD2009-00064 and CPAN CSD2007-00042, the LSC Consortium, and the Gobierno de Aragón and the European Social Fund (Group in Nuclear and Astroparticle Physics). Authors would like to acknowledge the use of Servicio General de Apoyo a la Investigación-SAI, Universidad de Zaragoza and technical support from LSC and GIFNA staff.

## References

[1] J. Amaré et al., *Performance of ANAIS-112 experiment after the first year of data taking*, *Eur. Phys. J. C* **79** (2019) 228 [1812.01472].

[2] J. Amaré et al., *Annual Modulation Results from Three Years Exposure of ANAIS-112*, *Phys. Rev. D* **103** (2021) 102005 [2103.01175].

[3] R. Bernabei et al., *Further results from DAMA/LIBRA-phase2 and perspectives*, *Nucl. Phys. Atom. Energy* **22** (2021) 329.

[4] J. Billard et al., *Direct detection of dark matter—APPEC committee report\**, *Rept. Prog. Phys.* **85** (2022) 056201 [2104.07634].

[5] J. Amare et al., *Analysis of backgrounds for the ANAIS-112 dark matter experiment*, *Eur. Phys. J. C* **79** (2019) 412 [1812.01377].

[6] I. Coarasa et al., *Improving ANAIS-112 sensitivity to DAMA/LIBRA signal with machine learning techniques*, *JCAP* **11** (2022) 048 ; erratum *JCAP* **06** (2023) E01 [2209.14113].

[7] T. Pardo et al., *Neutron calibrations in dark matter searches: the ANAIS-112 case*, in these proceedings .

[8] I. Coarasa et al., *ANAIS-112 sensitivity in the search for dark matter annual modulation*, *Eur. Phys. J. C* **79** (2019) 233 [1812.02000].